\documentclass[twocolumn,secnumarabic,amssymb, nobibnotes, aps, prl, superscriptaddress]{revtex4-1}
\usepackage{graphicx}                    
\usepackage{amssymb}                    
\usepackage{color}                       
\usepackage{subfigure}
\usepackage{url}

\begin{document}

\title{Anisotropy of Imbalanced Alfv\'{e}nic Turbulence in Fast Solar Wind}%

\author{R. T. Wicks}
\email{r.wicks@imperial.ac.uk}
\affiliation{Space and Atmospheric Physics Group, Imperial College London, London, SW7 2AZ, UK}
\author{T. S. Horbury}
\affiliation{Space and Atmospheric Physics Group, Imperial College London, London, SW7 2AZ, UK}
\author{C. H. K. Chen}
\affiliation{Space and Atmospheric Physics Group, Imperial College London, London, SW7 2AZ, UK}
\author{A. A. Schekochihin}%
\affiliation{Rudolf Peierls Centre for Theoretical Physics, University of Oxford, Oxford, OX1 3NP, UK.}

\begin{abstract}
We present the first measurement of the scale-dependent power anisotropy of Elsasser variables in imbalanced fast solar wind turbulence. The dominant Elsasser mode is isotropic at lower spacecraft frequencies but becomes increasingly anisotropic at higher frequencies. The sub-dominant mode is anisotropic throughout. There are two distinct subranges exhibiting different scalings within what is normally considered the inertial range. The low Alfv\'{e}n ratio and the different scaling of the Elsasser modes suggests an interpretation of the observed discrepancy between the velocity and magnetic field scalings, the total energy is dominated by the latter. These results do not appear to be fully explained by any of the current theories of incompressible imbalanced MHD turbulence.
\end{abstract}
\maketitle
\textit{Introduction.} The solar wind is an excellent plasma turbulence laboratory. Alfv\'{e}nic fluctuations in the fast solar wind (mean velocity $\gtrsim 600$ km/s) are well described by the incompressible MHD equations despite the collisionless nature of the plasma \cite{Schekochihin09}. The MHD turbulent cascade transports energy from large scales to smaller scales \cite[e.g.][]{Goldstein95} until it reaches the ion gyroscale, below which another type of turbulence carries energy to yet smaller scales \cite{Alexandrova, Chen10, Sahraoui}. In the fast wind, the turbulence is imbalanced: there is more power in Alfv\'{e}nic fluctuations traveling away from the Sun than toward it \cite{Tu89, Tu90}. There is evidence in the slow solar wind \cite{Lucek98} and from numerical simulations \cite{Perez09} that balanced turbulence is made up of locally imbalanced regions, so understanding imbalanced turbulence is probably essential for understanding MHD turbulence in general.
\par
A key property of plasma turbulence is anisotropy caused by the magnetic field. Even if the field is not strong enough to dominate the thermal pressure, its presence makes fluctuations scale differently in the field-perpendicular direction than in the field-parallel one, with larger power in fluctuations that vary across the field. This anisotropy is poorly understood: there is a relative dearth of observational data and an abundance of mutually contradictory theories. What currently appears to be the most compelling theory of the anisotropic Alfv\'{e}nic cascade is based on the assumption of `critical balance' \cite{GS95}, to which an assumption of `dynamic alignment' can be added \cite{Boldyrev06}. First posited for balanced cascades \cite{GS95} and later extended to imbalanced ones \cite{Lithwick07}, the critical balance conjecture states that the linear and nonlinear timescales are comparable and predicts anisotropic scalings of the fluctuation spectra: $E(k_{\perp}) \propto k_{\perp}^{-5/3}$ and $E(k_{||}) \propto k_{||}^{-2}$. The dynamic alignment conjecture \cite{Boldyrev06} states additionally that the polarizations of magnetic and velocity fluctuations align as the energy moves to smaller scales, which causes adjustment of the scalings to $E(k_{\perp}) \propto k_{\perp}^{-3/2}$ and $E(k_{||}) \propto k_{||}^{-2}$. 
\par
Recent numerical studies have shown a range of seemingly contradictory behaviors \cite[e.g.][]{Mason06, Beresnyak09, Perez09}, which were argued to agree, or disagree, with a number of conflicting theories \cite{Boldyrev06, Lithwick07, Beresnyak08, Chandran08, Podesta10}. So far, observational studies of anisotropy have focused on the magnetic field \citep{Horbury08, Podesta09, Luo10, Wicks10, Chen10b}, but a complete analysis of the Alfv\'{e}nic turbulence must include the velocity field. The Elsasser fields \cite{Elsasser}, $\textbf{Z}^{\pm} = \textbf{V} \pm \textbf{B} / \sqrt{4\pi\rho_0}$, where $\textbf{V}$ and $\textbf{B}$ are the velocity and magnetic fields respectively, and $\rho_0$ is the average mass density, describe oppositely propagating (at local Alfv\'{e}n speed) finite-amplitude solutions of the incompressible MHD equations. There is a good physical case for treating these as the primary fields that make up the Alfv\'{e}nic turbulence, which can be thought of as the result of their interactions \cite{Kraichnan65}. This Letter presents the first \textit{in-situ} solar wind observation designed to measure anisotropy of Elsasser variables as well as the anisotropy of the magnetic and velocity fields. The results do not appear to be in quantitative agreement with any of the current theories.
\par
\textit{Data analysis.} We use data obtained by the WIND spacecraft at 3-second spin resolution. Magnetic field $(\textbf{B})$ is provided by the MFI instrument, velocity $(\textbf{V})$ and density $(\rho)$ from the 3DP instrument. Data are taken from a long-duration fast stream from days 13 to 20 of 2008, in which the solar wind speed remained above 550 km/s for the entire 7-day interval and had an ion plasma beta of $1.1$. Compressive fluctuations were an order of magnitude weaker in power than incompressible, so the magnetic fluctuations are dominated by the component perpendicular to the mean magnetic field. We denote the (dominant) Alfv\'{e}nic fluctuations traveling away from the Sun by $\textbf{Z}^{+}$ for easier comparison to previous work \cite[e.g.][]{Tu90}. Other, shorter fast-stream periods show similar results but have larger errors due to insufficient statistics in smaller data sets.
\par
\begin{figure*}
\subfigure
{
\includegraphics[width=\columnwidth]{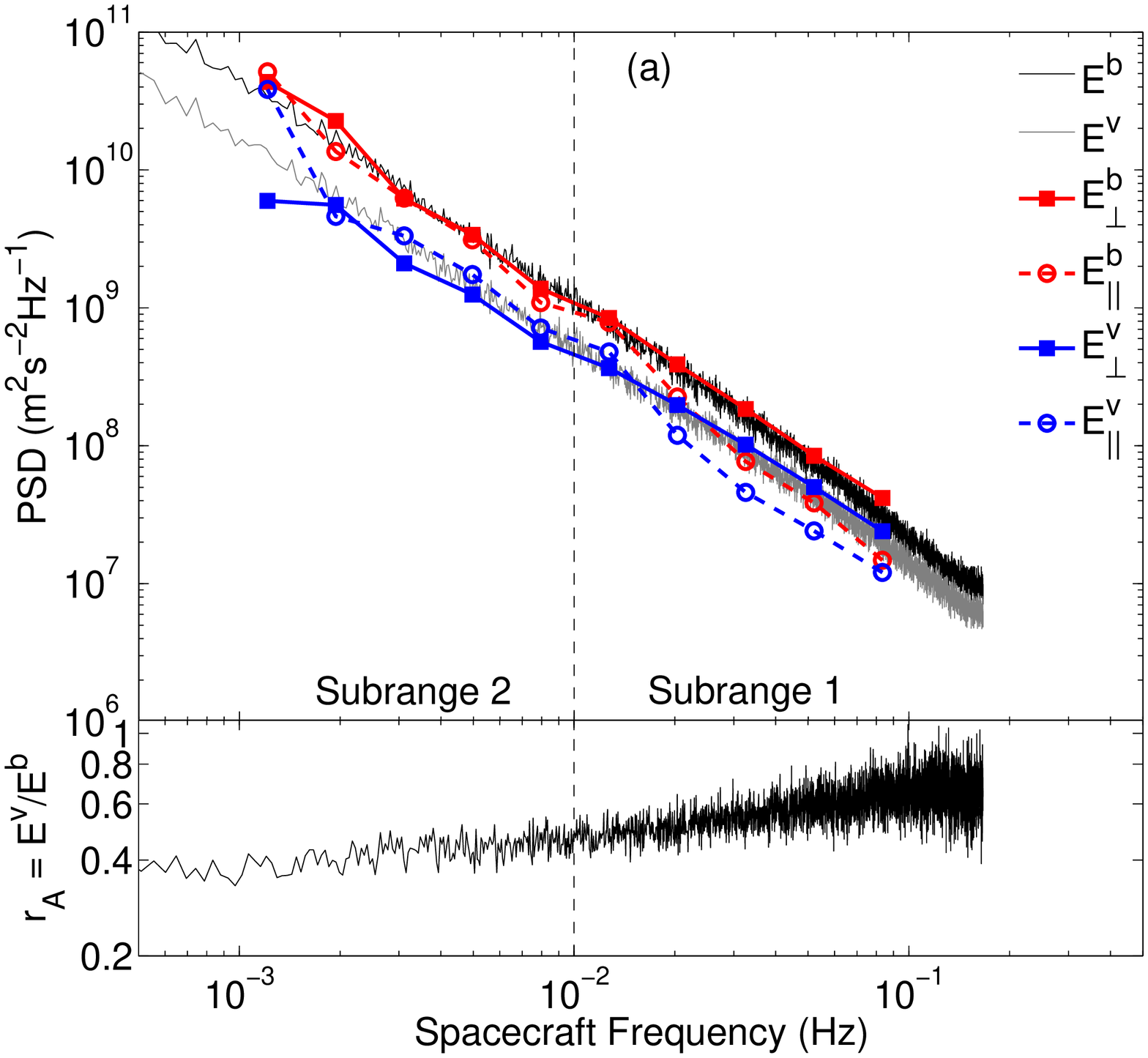}%
\label{fig:1subfiga}
}
\subfigure
{
\includegraphics[width=\columnwidth]{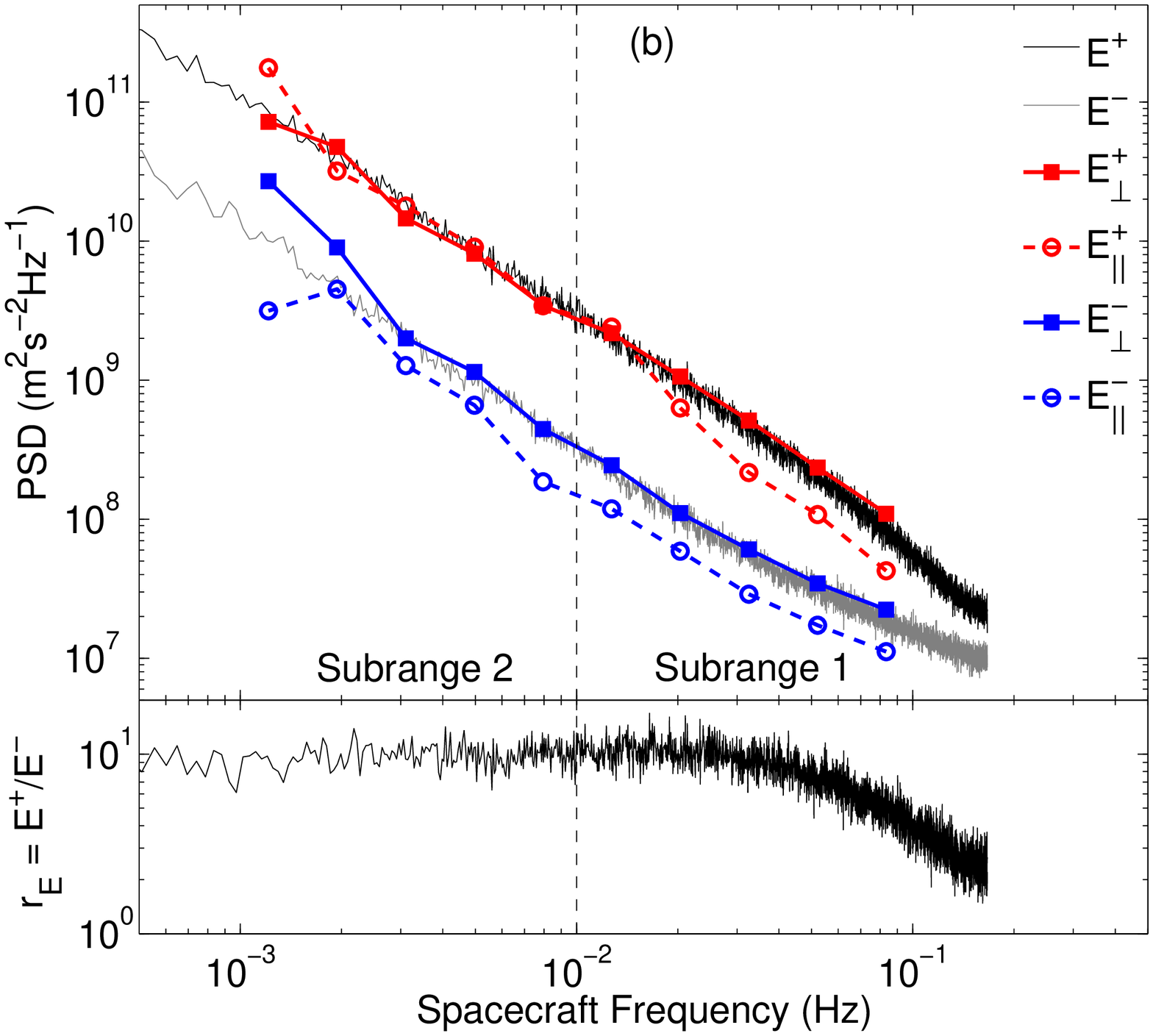}%
\label{fig:1subfigb}%
}
\caption{Power spectra from WIND data between days 13 and 20 of 2008. (a) Trace of the Fourier and wavelet power spectra $E^v$ (gray line and blue symbols) and $E^b$ (black line and red symbols). The bottom panel shows the Alfv\'{e}n ratio $r_A = E^v/E^b$. (b) Trace of the Fourier and wavelet power spectra of $E^+$ (black line and red symbols) and $E^-$ (gray line and blue symbols). The bottom panel shows $r_E = E^+/E^-$, the high frequency decrease in which may be due to quantization as discussed in the text.}
\label{fig:1}
\end{figure*}
We use Morlet wavelets \cite{TorrenceCompo} to measure the power in fluctuations of $\textbf{B}$ and $\textbf{V}$ as a function of time and scale \cite{Horbury08, Podesta09, Wicks10}. The time resolution of the Morlet wavelet is provided by a Gaussian envelope function, the width of which changes with scale. This width is used as the scale over which we calculate the average density $(\rho_0)$ and the average angle $\theta_B$ between the magnetic field and the measurement direction (radial). The power in the Elsasser fields is then calculated by combining the wavelet coefficients of $\textbf{V}$ and $\textbf{B}$. In this way we calculate the trace power spectra, of $\textbf{B}$, $\textbf{V}$, $\textbf{Z}^+$ and $\textbf{Z}^-$ as a function of spacecraft frequency $f$ (proportional to scale under Taylor's hypothesis) and the angle $\theta_B$. The angle bins are $10^{\circ}$ wide and the frequency bins are logarithmically separated by a factor of $1.6$. We stress that anisotropic power spectra calculated in this way do not represent power in individual components of the vector fields but rather the trace power averaged over instances when the local mean magnetic field points in the direction given by angle $\theta_B$ with respect to the solar wind velocity.
\par
The trace power spectra of $\textbf{V}$ and $\textbf{B}/\sqrt{4\pi\rho_0}$ are denoted by $E^{v}$ and $E^{b}$ respectively, similarly the power spectra of $\textbf{Z}^\pm$ are $E^\pm$. Subscripts $\perp$ and $||$ denote the trace spectra corresponding to fluctuations that vary perpendicularly ($80^{\circ} < \theta_B < 90^{\circ}$) and parallel ($0^{\circ} < \theta_B < 10^{\circ}$) to the local magnetic field. We refer to the measured change in power with $\theta_B$ at a fixed frequency as `power-level anisotropy' and to the change in the measured scaling of the power spectrum with $\theta_B$ as `spectral-index anisotropy'.
\par
\textit{Spectra.} We plot $E^{v}$ and $E^{b}$ in the top panel of Fig.\ref{fig:1subfiga}. The Fourier power of $E^b$ is larger than $E^v$ at all scales, as can be seen in the bottom panel which shows the Alfv\'{e}n ratio $r_A = E^v/E^b$; at low frequencies, the magnetic field dominates with $r_A \approx 0.4$, but as frequency increases, $r_A$ approaches unity. The wavelet power measured parallel and perpendicular to the magnetic field is also plotted in Fig.\ref{fig:1subfiga}. The magnetic field has equal power in both directions for $f \lesssim 10^{-2}$ Hz but is anisotropic at higher frequencies, with less power in the parallel direction. This is similar to the known result of magnetic field anisotropy starting at the outer scale of turbulence reported for Ulysses data \citep{Luo10, Wicks10}, but the transition from isotropy to anisotropy occurs at a significantly higher frequency than what one would typically call the outer scale in such observations ($10^{-2}$ Hz rather than $10^{-4}$ Hz).
\par
\begin{figure*}
\subfigure
{
\includegraphics[width=\columnwidth]{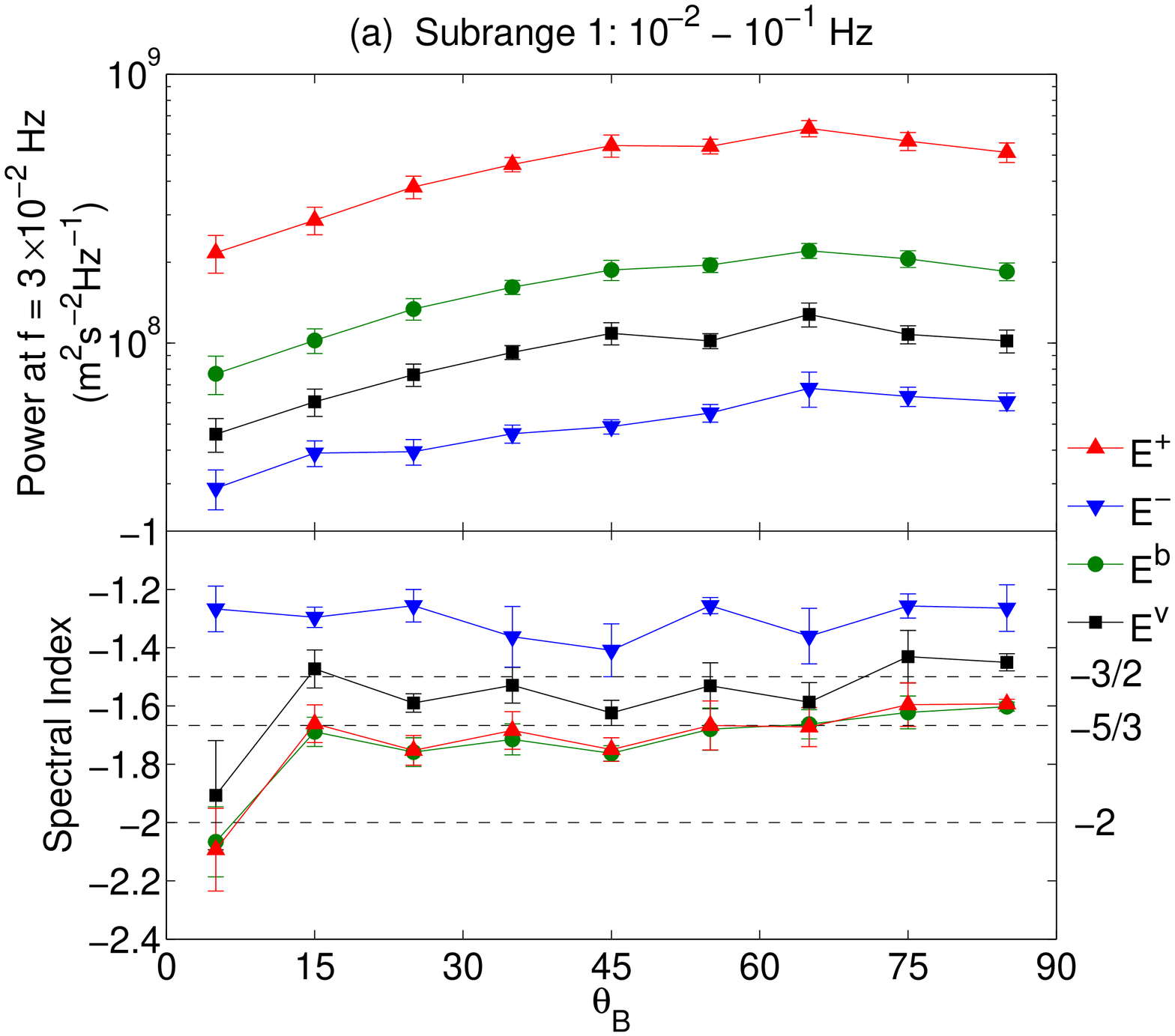}%
\label{fig:2subfiga}%
}
\subfigure
{
\includegraphics[width=\columnwidth]{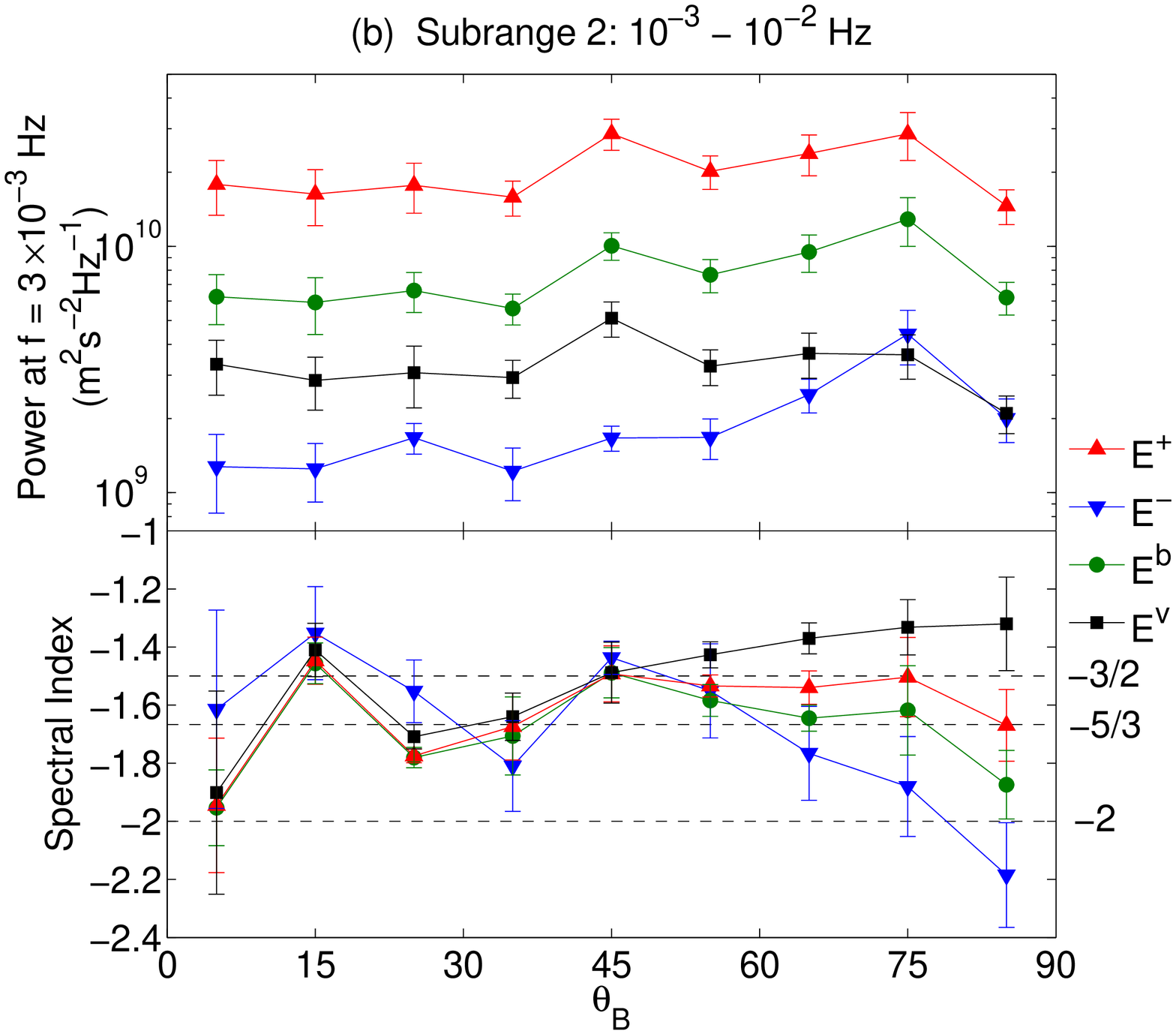}%
\label{fig:2subfigb}%
}
\caption{Power and spectral index anisotropy of the two subranges of the power spectra in Fig. 1. (a) Subrange 1 ($10^{-1}$ and $10^{-2}$ Hz): variation of power (top panel) at $3.26 \times 10^{-2}$ Hz and spectral index (bottom panel) with $\theta_B$. (b) Subrange 2 ($10^{-2}$ and $10^{-3}$ Hz): variation of power (top panel) at $3.1\times10^{-3}$ Hz and spectral index (bottom panel) with $\theta_B$.}
\label{fig:2}%
\end{figure*}
The wavelet velocity spectra have lower power than $E^b$ but behave in a similar manner. As the magnetic field becomes anisotropic at frequencies $\gtrsim 10^{-2}$ Hz, so does the velocity, with the parallel power decreasing and the perpendicular power dominating. Thus, $E^b$ and $E^v$ track each other, suggesting a common source. Error bars are not plotted here to keep the figures clear; errors on individual wavelet power observations grow as frequency decreases, being the same size as the circle and square markers at the highest frequency and increasing in size to about $50\%$ of the wavelet value at the lowest frequencies ($\sim 10^{-3}$ Hz). 
\par
Fig.\ref{fig:1subfigb} shows the Fourier power spectra $E^+$ and $E^-$ (top panel) calculated for the same period as Fig.\ref{fig:1subfiga}. The Elsasser variables show imbalance, with $E^+$ dominant over $E^-$ at all frequencies. This is quantified in the bottom panel which shows $r_E = E^+/E^-$, at all frequencies for this period $r_E > 0$; for purely anti-sunward Alfv\'{e}n waves we expect $r_E = \infty$. Fig.\ref{fig:1subfigb} also shows the Elsasser power spectra parallel and perpendicular to the local mean magnetic field. The dominant ($\textbf{Z}^+$) modes are isotropic in power at frequencies lower than $10^{-2}$ Hz, but grow increasingly anisotropic at higher frequencies. This behavior is similar to that for the magnetic field in Fig.\ref{fig:1subfiga}, which is expected since $r_A < 1$. The weaker ($\textbf{Z}^-$) mode behaves differently, with power-level anisotropy measured at all scales. 
\par
Thus, we have found that what is usually thought of as the `inertial range' of solar wind turbulence in fact consists of two distinct subranges, each about a decade wide in this data interval: the higher-frequency Subrange 1 $(\sim 10^{-2}$ - $10^{-1}$ Hz$)$ and the lower-frequency Subrange 2 $(\sim 10^{-3}$ - $10^{-2}$ Hz$)$. The scalings and anisotropy are clearly different in these two subranges. We now proceed to quantify these differences in terms of power-level and spectral-index anisotropy.
\par
\textit{Scaling and anisotropy in Subrange 1.} The top panel of Fig.\ref{fig:2subfiga} shows the power at $f = 3.26 \times 10^{-2}$ Hz, plotted against $\theta_B$ for each of the four fields. The bottom panel shows the spectral indices measured over Subrange 1. There is power-level anisotropy in all 4 variables with the ratios $E^b_{\perp}/E^b_{||} = 2.4 \pm 0.2$, $E^v_{\perp}/E^v_{||} = 2.2 \pm 0.2$, $E^+_{\perp}/E^+_{||} = 2.4 \pm 0.2$, and $E^-_{\perp}/E^-_{||} = 2.0 \pm 0.2$, although these ratios change with scale for $E^b$, $E^v$ and $E^+$ due to their spectral-index anisotropy. Errors in power are the standard deviation of the data contributing to each mean, and errors in the spectral index are calculated from the least-squares fit of a straight line to the wavelet power spectra on a log-log scale.
\par
The bottom panel of Fig.\ref{fig:2subfiga} attests that the spectral index of $E^+$ is almost identical to $E^b$, with steeper spectra at smaller $\theta_B$ and the spectral index tending to $-2$ at $\theta_B = 0$, consistent with standard theories based on critical balance \cite{GS95, Boldyrev06}. For larger angles $(\theta_B > 15^{\circ})$, $E^+$ and $E^b$ have spectral indices around $-5/3$. The $E^v$ spectral indices are slightly shallower than the $E^b$ and $E^+$ values throughout and tend towards $-3/2$ for $\theta_B > 15^{\circ}$, consistent with the shallower gradient of the $E^v$ Fourier spectrum compared to $E^b$ (see Fig.\ref{fig:1subfiga} and \cite[e.g.][]{Podesta07}). $E^-$ is clearly different from the other fields: it shows no obvious spectral-index anisotropy, which also has much shallower values around $-1.3$. Thus, the sub-dominant Elsasser field has power-level anisotropy but its anisotropy does not change with scale. 
\par
There is a caveat concerning the $E^-$ results, however. The WIND 3DP velocity data are subject to quantization; the velocity observations are digitized in such a way that the velocity appears measured in discrete steps at small scales rather than as a smoothly varying time series. A careful analysis shows that the effect on $E^v$ and its anisotropy is small due to the low power of the quantization; however the Elsasser variables are more strongly affected. The quantization acts to decorrelate the magnetic and velocity fluctuations at frequencies higher than $f \sim 3\times10^{-2}$ Hz, increasing the power in the weaker Elsasser variable with a corresponding slight decrease in the dominant mode. Only the effect on the weaker Elsasser mode is significant relative to the errors here, and as such the results for $f > 3\times10^{-2}$ Hz represent a minimum bound on the power-level anisotropy of the weaker Elsasser variable. The quantization is unlikely to have affected the power-level anisotropy in Fig.\ref{fig:2subfiga} since we have chosen a scale where the noise effect is small. The $E^-$ spectral indices are shallower than they should be over this frequency range and the quantization may remove any spectral-index anisotropy. The power-level anisotropy does not change significantly at lower frequencies, however, confirming the lack of spectral-index anisotropy in general, if not the exact values of the spectral indices in this subrange.
\par
\textit{Scaling and anisotropy in Subrange 2.} The top panel of Fig.\ref{fig:2subfigb} shows the power of each variable plotted vs. $\theta_B$ at $f = 3.1\times10^{-3}$ Hz. $E^+$, $E^b$ and $E^v$ are approximately flat, exhibiting no systematic power-level anisotropy. $E^-$, however, retains a level of anisotropy similar to that seen in Subrange 1, with greater power at larger angles than in the parallel direction. The power-level anisotropy ratios are $E^b_{\perp}/E^b_{||} = 1.1 \pm 0.3$, $E^v_{\perp}/E^v_{||} = 0.7 \pm 0.3$, $E^+_{\perp}/E^+_{||} = 0.9 \pm 0.3$, and $E^-_{\perp}/E^-_{||} = 1.7 \pm 0.4$. The bottom panel shows the spectral indices plotted against $\theta_B$. Subject to larger errors than in subrange 1, the spectral indices of $E^b$ and $E^+$ do not show any measurable change with angle, and hover around the $-5/3$ and $-3/2$ values. $E^v$ shows similar spectral indices to $E^b$ and $E^+$ for $\theta_B < 45^{\circ}$ but its spectral index gets shallower at larger angles, increasing to around $-1.3$ in the perpendicular direction. The spectral index of $E^-$ shows an entirely different trend: it is flatter at around $-1.5$ in the parallel direction, becoming steeper at larger angles until it is less than $-2$ at $90^{\circ}$. This is in agreement with the result that the sub-dominant Elsasser mode has a steeper spectrum than the dominant one at lower frequencies \citep{Tu89, Tu90}.
\par
\textit{Conclusions and discussion.} We have presented the first observations of power-level and spectral-index anisotropy of Elsasser variables in imbalanced fast solar wind turbulence. The two remarkable conclusions of this study are as follows. (i) What is usually thought of as a simple `inertial range' is in fact split into two subranges. In the data used here the split is at $f \sim 10^{-2}$ Hz, where both the scaling behavior and the anisotropy of all fields change. (ii) While \textbf{V}, \textbf{B} and $\textbf{Z}^+$ have similar anisotropy (very little in the lower-frequency subrange, increasing with frequency in the higher-frequency subrange), the sub-dominant Elsasser field is completely different: it has power-level anisotropy but in a scale-independent way at higher frequencies, and scales steeply in the perpendicular direction at lower frequencies. $\textbf{Z}^+$ and \textbf{B} behave very similarly at all scales and the anisotropy of $\textbf{B}$ and $\textbf{V}$ appear related; the change from isotropy to anisotropy occurs at the same frequency in both, although the spectral indices are shallower for the velocity than for the magnetic field.
\par
We believe that our results shed some light on the theoretically puzzling observations that suggest different spectral scalings of $\textbf{V}$ and $\textbf{B}$, $-3/2$ for the former and $-5/3$ for the latter \cite{Podesta07}, a behavior never observed in simulations or envisioned in theories. In the solar wind, the Alfv\'{e}n ratio is significantly below 1 resulting in $\textbf{B}$ and $\textbf{Z}^+$ scaling in the same way. Since $\textbf{Z}^-$ has a shallower scaling, the $\textbf{V}$ scaling lies between these two behaviors and the closeness of the spectral exponent to $-3/2$ might, in fact, be a coincidence. The dominant contribution to the energy comes from $\textbf{Z}^+$ (and so $\textbf{B}$), which has a robust $-5/3$ scaling.
\par
Although these results in their entirety are not precisely reproduced by any of the extant simulations or theory, there are some points of similarity. The MHD simulations of \cite{Beresnyak09} show different scaling and anisotropy in the two Elsasser variables for imbalanced turbulence (the dominant $\textbf{Z}^+$ scales in a critically balanced way and the weaker $\textbf{Z}^-$ scales differently, although the scaling of $\textbf{Z}^-$ does not agree with our findings). The theory of \cite{Chandran08} also predicts different scaling for Elsasser variables, with the concept of `pinning' of $E^+$ and $E^-$ to the same power at the dissipation scale. In particular, the results from a mixture of weak and strong turbulence in \cite{Chandran08} are qualitatively similar to the results here, with a break in the middle of what appears to be an inertial range and a change from steeper to shallower scaling of $\textbf{Z}^-$, although again the anisotropy of $\textbf{Z}^-$ does not agree with our results.
\par
This work was supported by STFC and the Leverhulme Trust Network for Magnetized Plasma Turbulence. WIND data were obtained from the NSSDC website.

\end{document}